| | |
|---|---|
| **Title** | **Synthesis of Membrane-Electrode Assembly for Fuel Cells by Means of (Sub)-Atmospheric Plasma Processes** |
| **Authors** | Delphine Merche, Thierry Dufour, Julie Hubert, Claude Poleunis, Sami Yunus, Arnaud Delcorte, Patrick Bertrand, François Reniers* |
| **Affiliations** | Service de Chimie Analytique et Chimie des Interfaces (CHANI), CP255, Université Libre de Bruxelles (ULB), Bv du Triomphe, 2, B-1050, Brussels, Belgium<br>Institut de la Matie`re Condense´e et des Nanosciences (IMCN), Bio and Soft Matter (BSMA), Universite´ Catholique de Louvain (UCL), Croix du Sud 1, box L07.04.01, B-1348 Louvain la-Neuve, Belgium |
| **Ref.** | Plasma Processes & Polymers, 2012, Vol. 9, Issue 11-12, 1144-1153 |
| **DOI** | http://dx.doi.org/10.1002/ppap.201100208 |
| **Abstract** | An easy procedure to build up membrane-electrode assemblies for applications dedicated to miniaturized PEMFC using $H_2$ or $CH_3OH$ by a two-steps atmospheric plasma process is reported. Firstly, catalyst nanoparticles are grafted on carbon substrates by spraying a Pt colloidal solution in the post-discharge of an RF atmospheric plasma torch. In the second step, the resulting decorated electrodes are covered by plasma synthesized polymeric membranes in the discharge of a DBD. The sulfonated polystyrene membranes are synthesized by injecting simultaneously styrene and trifluoromethanesulfonic acid monomers, in the presence of a carrier gas (Ar or He). The membranes are chemically characterized by XPS, ToF-SIMS, and FTIR (IRRAS) and their deposition rate is investigated by SEM. |

# 1. Introduction

Fuel cells are subject to a widespread enthusiasm as they constitute an alternative to conventional batteries and accumulators, and allow us to bring complementary solutions to fossil energies.[1] Many studies have been conducted for several years in order to optimize the triple contact of the fuel cells (arrival of the gases/ionomer/catalyst) by improving the different interfaces and to reduce the price of the different elements constituting the cell. Plasma processes for the synthesis of the membranes and electrodes are a major asset because the microstructural and chemical properties of the plasma-membranes (chemical composition, structure, morphology, thickness, etc.) and the plasma-electrodes are closely linked to the deposition conditions. Moreover, these environmental-friendly processes could reduce the amount of noble metal for the electrodes, while preserving the same electrical performances. Also, they allow the synthesis of thin ion-exchange membranes in one single step, contrary to liquid phase-based methods [2–4] which require multiple steps, organic solvents, reactants, and long reaction times. This synthesis is especially adapted to microfuel cells supplied in methanol or hydrogen dedicated to the electronic market (notebooks, cameras, mobile phones, small robots, etc.). Moreover, the strong adherence of the plasma-membranes onto electrodes constitutes an advantage for the intimate contact between the catalyst/membrane/arrival of the reactants where the redox reactions take place in the cell.

The studies relative to the synthesis by ‘‘Plasma Enhanced Chemical Vapor Deposition’’ (PECVD) of thin ionic membranes from a polymerizable monomer (fluorocarbon or hydrocarbon backbone monomer) mixed with sulfonation agents (trifluoromethanesulfonic acid, chlorosulfonic acid, sulfur dioxide, etc.) are generally carried out at low pressure in order to stabilize the species and the plasma discharge. [5–13]

In our work, plasma-polymerized sulfonated polystyrene (pp-sulfonated PS) was synthesized in one single step from a mixture of styrene and trifluoromethanesulfonic acid ($CF_3$-$SO_3H$) in the discharge of a dielectric barrier discharge (DBD) under sub-atmospheric pressure.

In addition to the PECVD of the membranes on usual substrates (as silicon) for chemical characterization, membrane-electrode assemblies were performed by a two-steps plasma process.

Firstly, catalyst nanoparticles were grafted on carbon substrates by spraying a Pt colloidal solution in the post-discharge of an RF atmospheric plasma torch (according to previous papers and patent from our team).[14–18] This fast and easy method allows the surface activation and the surface decoration in only one step. The activation of carbon substrates due to plasma exposure is usually based on the grafting of oxidized species but also on the formation of structural surface defects due to ion and electron bombardment, which can, in turn, be functionalized.[17] This step is often carried out by means of wet chemistry techniques.[19] In other cases, the activation of the surface and the grafting of metal particles are combined with techniques requiring a high vacuum system (ion gun, low pressure plasma, and/or thermal evaporation).[20] Secondly, plasma-polymer membranes were deposited onto the Pt-plasma-decorated electrodes directly in the DBD, operating at sub-atmospheric pressure, as described previously.

As the two-steps process could be easily combined in one continuous procedure and as it leads to a strong contact between the membrane and the catalytic phase, the perspective to constitute a financially viable alternative compared to the industrial synthesis of the Nafion is appealing. Moreover, the Nafion presents disadvantages as permeability of the fuel, low temperature limit and a relatively high thickness (a hundred of mm) which are locks for the use of the commercial membrane in miniature devices.[1,21,22]

Working under (sub)-atmospheric pressure presents lots of advantages: it avoids the constraints linked to a costly high-vacuum system, and therefore enables an easy implementation of the process in a continuous production line. However, a major drawback due to the high pressure is the very small mean free path of the reactive species, making the reaction mechanisms much more difficult to understand, and therefore to control. The pp-sulfonated PS obtained by plasma copolymerization were chemically characterized by X-ray photoelectron spectroscopy (XPS), infrared reflection absorption spectroscopy (IRRAS), and by dynamic and static time-of-flight secondary ion mass spectrometry (ToF-SIMS). The deposition rate was studied by scanning electron microscopy (SEM). The thermal stability and the ability to exchange ions (after soaking in an electrolyte solution) were investigated by XPS. The carbon substrates decorated by Pt nanoparticles were characterized by XPS, static ToF-SIMS, and SEM. Dynamic ToF-SIMS and SEM were used to observe the different interfaces of the pp-sulfonated PS deposited onto glassy carbon decorated by Pt nanoparticles.

# 2. Experimental Section

## 2.1. Materials

The liquid styrene (Fluka, purity: 99.5%) and the liquid trifluoromethane sulfonic acid precursors (Sigma–Aldrich, purity >98%) were used as received for the plasma polymerization, without any further purification. This acid being very hygroscopic, it requires a closed system with a controlled atmosphere. The Pt nanoparticles were provided by Plasmachem. They were sterically stabilized by polyvinylpyrrolidone (PVP) and showed a size distribution of 3–4nm in diameter. Once diluted in methanol, the resulting solution was used for the first step of the process.
Silicon wafers (100) from Compart Technology Ltd. were selected as flat substrates for surface characterizations and for the deposition rates study of the pp-sulfonated PS. Glassy carbon (non-porous) and Carbon Toray Paper (porous) were used for the grafting of Pt catalyst and then for the membrane synthesis (membrane-electrode assembly). Glassy carbon (Sigradur K platen) was chosen to this end as a model for the depth profiling by dynamic SIMS. The substrates were cleaned with methanol and isooctane. The atmospheric plasma torch used for the grafting of the metallic nanoparticles was sustained by Ar (''Air liquide'', purity: 99.999%), mixed with oxygen (''Air liquide'', purity: 99.995%). The DBD for the plasma-polymer synthesis was sustained either by He (''Air liquide'', purity: 99.999%) or by Ar (''Air liquide'', purity: 99.999%).

## 2.2. Nanoparticles Grafting

The catalyst nanoparticles were deposited on carbon gas diffusion layers (Carbon Toray Paper) and on glassy carbon by spraying the Pt colloidal solution in the post-discharge of an RF atmospheric plasma torch. The plasma torch (Atomflo250DSurfX Technologies) consists of two closely spaced circular aluminum electrodes perforated to flush the process gas (Ar) at a flow rate of 30 L.min$^{-1}$. To improve the grafting process on the carbon substrates, a tiny flow rate of oxygen (20mL.min$^{-1}$) was mixed with the carrier gas, as already highlighted for the grafting of Au, Rh, and Pt nanoparticles in the Nano2hybrids project. [15]

**Fig. 1**

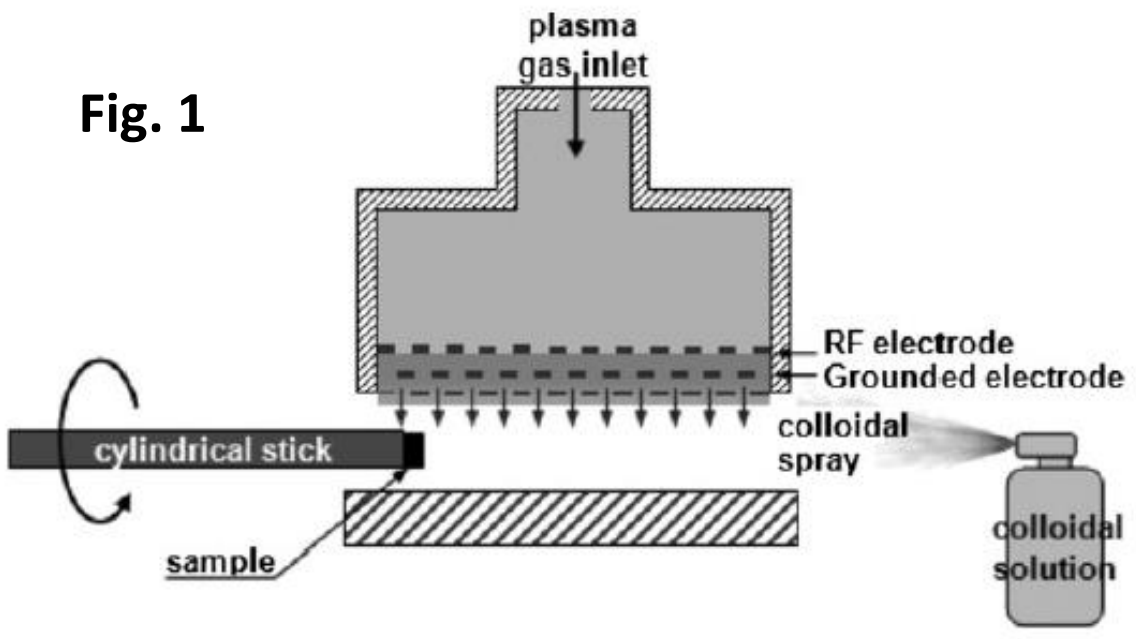

***Figure 1. Experimental setup dedicated to the grafting of catalytic nanoparticles, by using an RF plasma torch.***

***Figure 2. Experimental setup dedicated to the pp-sulfonated PS on Si wafers or on carbon substrates decorated with Pt nanoparticles (NPT) by using a home-built DBD. The inset shows a schematic representation of pp-sulfonated PS synthesized on glassy carbon decorated with Pt nanoparticles.***

**Fig. 2**

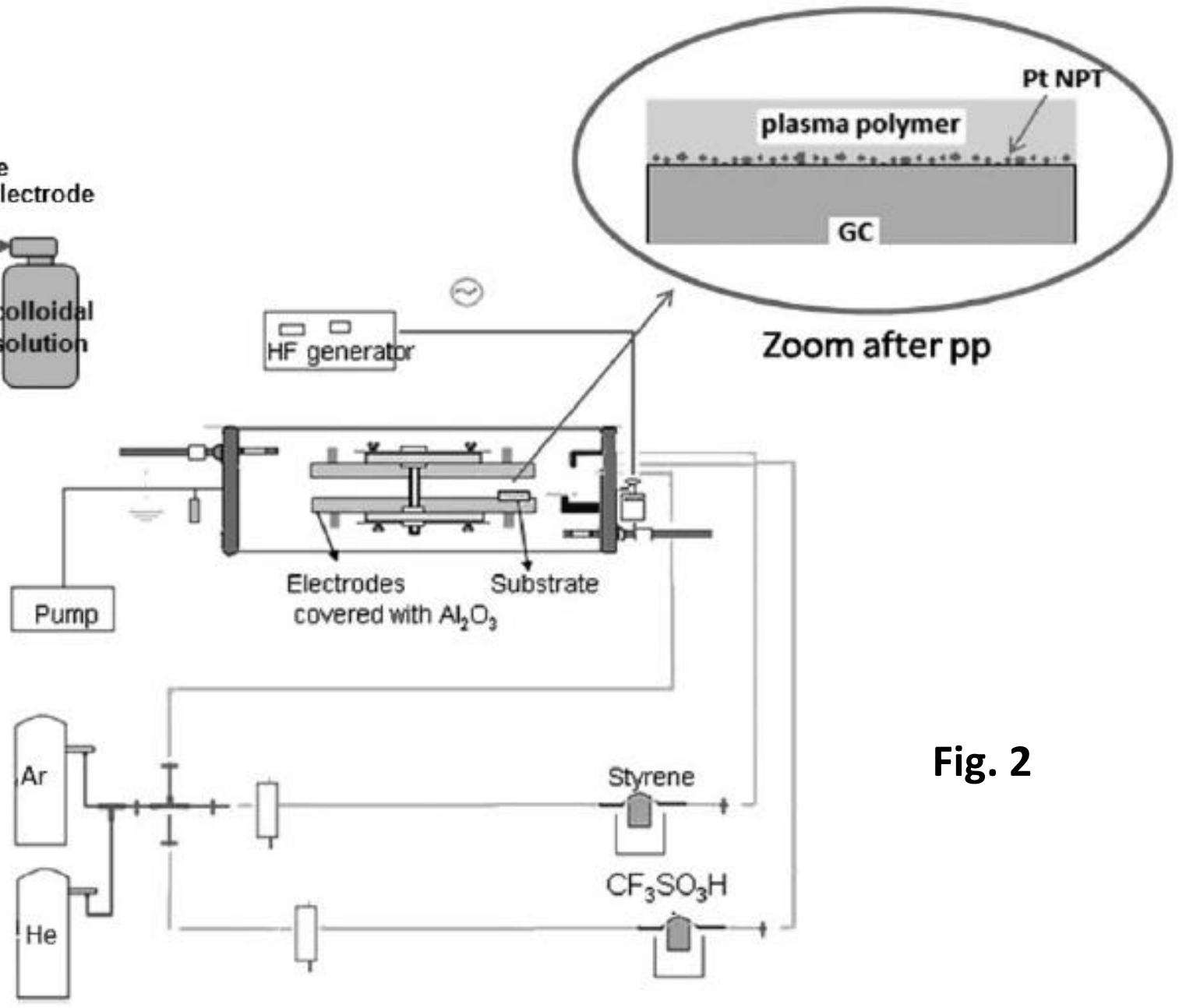

We have carried out several configurations (variation of the angle between the substrate and the gas flow direction, injection of the colloidal solution out of the post-discharge, etc.). The best configuration resulted from the best compromise between a high amount of nanoparticles grafted and a low oxidation of the Pt nanoparticles, as illustrated in Figure 1. The substrate, parallel to the gases flow direction, is fixed on a rotative cylindrical stick of 10mm in diameter and placed at 2mm from the plasma torch exit. The substrates were activated during 2min (30 L.min$^{-1}$ Ar, 20mL.min$^{-1}$ $O_2$ at 80W). Then the Pt colloidal solution was nebulized in the afterglow of the plasma torch during 3min (20 sprays of 0.17mL). After the 3min of Pt grafting, a termination step is operated under the same conditions without any spraying of platinum (during 2min).

## 2.3. Plasma Polymerization

The pp-sulfonated PS thin films were deposited in one single step, in a home-built DBD. The cylindrical glass chamber has already been detailed in previous papers. [23,24] The schematic diagram of the reactor is presented in Figure 2. Although the system also runs under atmospheric pressure, the results presented in this paper were obtained at a slightly reduced pressure ($2.10^4$ Pa) but still in the same order of magnitude as atmospheric pressure, in order to be able to increase the inter-electrode distance and to minimize the breakdown voltage (according to the Paschen Law). The chamber was pumped continuously during the plasma polymerization process (dynamic regime) to minimize the air contamination of the growing film. The liquid styrene monomer was kept at 313K and the temperature of the trifluoromethane sulfonic acid was ranging from 313 to 333 K.

Both monomers vapors were simultaneously introduced into the discharge with identical flow rates of Ar or He (1.5 $L.min^{-1}$ flowing through each liquid vessel). The frequency was fixed at 15 kHz and the voltage was varied from 1 to 2 kV. The silicon substrates (or the decorated carbon substrates) were placed between the electrodes, close to the arrival of the precursors (on the right end side of the 21cm long bottom electrode, at 9cm from its center, as shown in Figure 2). Deposition times ranged from 5 to 10 min. After the plasma polymer deposition on Si wafers or on carbon electrodes loaded with Pt catalyst, the chamber was purged in order to evacuate the reactive gas, then backfilled to atmospheric pressure using Ar or He. Finally, the samples were transferred to the analytical tools.

## 2.4. Characterization of the Samples

The pp-films were characterized by FTIR, XPS, and ToF-SIMS. The deposition rates were calculated by measuring the thickness (by cross-section) of the pp-sulfonated PS films by means of a Field Emission Gun Scanning Electron Microscope (FEG-SEM) Zeiss Leo 982. SEM was also used to investigate the morphology of the plasma-polymerized films (with the addition of a gold coating of about 5nm in thickness to make them conductive) and the deposited Pt. To evaluate the quality of the grafting, XPS analyses were performed on the carbon substrate decorated with the catalyst nanoparticles before and after ultrasonification in a methanol bath.

XPS analyses were carried out in a Physical Electronics PHI-5600 system equipped with a concentric spherical analyser and a 16 channeltron plate. Spectra were acquired with the Mg anode (1 253.6 eV) operating at 300 W. The typical area of analysis is around 800mm. The average surface concentrations (excluding hydrogen, which is not detected in XPS) on two identical samples at three different regions were computed on the Casa XPS software, after removal of a Shirley background line.

The sensitivity coefficients used are: S=0.35, C=0.205, N=0.38,O=0.63, F=1,andPt=3.50, instead of 1.75 (corresponding to the singlet Pt $4f_{7/2}$) as referred in the literature.[25] Indeed, in order to take into account all the platinum components and not to neglect the Pt (+4) and Pt (+2) components, we have considered the whole spectral envelope on the survey spectrum, corresponding to the spin-orbit doublet (Pt $4f_{7/2}$ and Pt $4f_{5/2}$). Therefore the compositions of the Pt decorated carbon substrates must be taken as indicative and are used only for comparison between the different plasma treatments (with/without oxygen, nature of the substrate, etc.).

Wide survey spectra were acquired at a 93.9 eV pass-energy, with a 5 scan accumulation (time/step: 50ms, eV/step: 0.8). The high resolution scans were acquired at a 23.5 eV pass-energy with an accumulation of 10 scans (time/step: 50ms, eV/step: 0.025). The high resolution scans of C1s, S2p, and Pt4f were decomposed into several peaks components, with the same FWHM for each, through convoluting Gaussian and Lorentzian profiles with respect to a 30:70 ratio. The components assignments and binding energies were based on the literature. [9,26,27]

Static positive and negative ToF-SIMS measurements were performed with an ‘‘IONTOF V’’ spectrometer (IONTOF GmbH, Münster, Germany) equipped with a reflectron analyser and controlled by the SURFACELAB 6.1 software. The sample was bombarded in static regime at two different regions by pulsed $Bi_3^{2+}$ (acceleration voltage=60 keV, primary ion current=0.01pA, total ion fluence=2.10$^{13}$ $Bi_3^{2+}$ cm$^{-2}$, beam diameter 0.3µm). The acquisition time was fixed to 1min and the scan was confined to a square of 500*500µm2. $Cs^+$ (acceleration voltage=1 keV, primary ion current=100nA, total ion fluence=1,7.10$^{-18}$ $Cs^+$.cm$^{-2}$, beam diameter = 0.3mm) was employed as primary source for the dynamic ToF-SIMS measurements (erosion of 100*100µm$^2$) on pp-sulfonated deposited on glassy carbon decorated with Pt nanoparticles.

# 3. Results and Discussion

## 3.1. pp-Sulfonated PS on Si Wafers

**Chemical Structure**

XPS is a very powerful technique to characterize thin polymer coatings.[28] Figure 3 presents the survey of pp sulfonated PS/DBD He and high resolution scans of S2p, C1s, O1s, and F1s. We have already shown that the synthesis of pp sulfonated PS revealed a relatively preserved structure of the acid precursor in the discharge near atmospheric pressure in our plasma conditions.[23] All the sulfur is present in the sulfonic form (S2p1/2 and S2p3/2 at 169 and 170 eV, respectively) and no sulfur at lower oxidation state (such as $RSO_2R'$ at 166 eV and RSOR’ at 164.5 eV) is incorporated into the films. These results differ from most of the deposits operated at low pressure, where the peak is often splitted.[5,8,9,22,29,30]

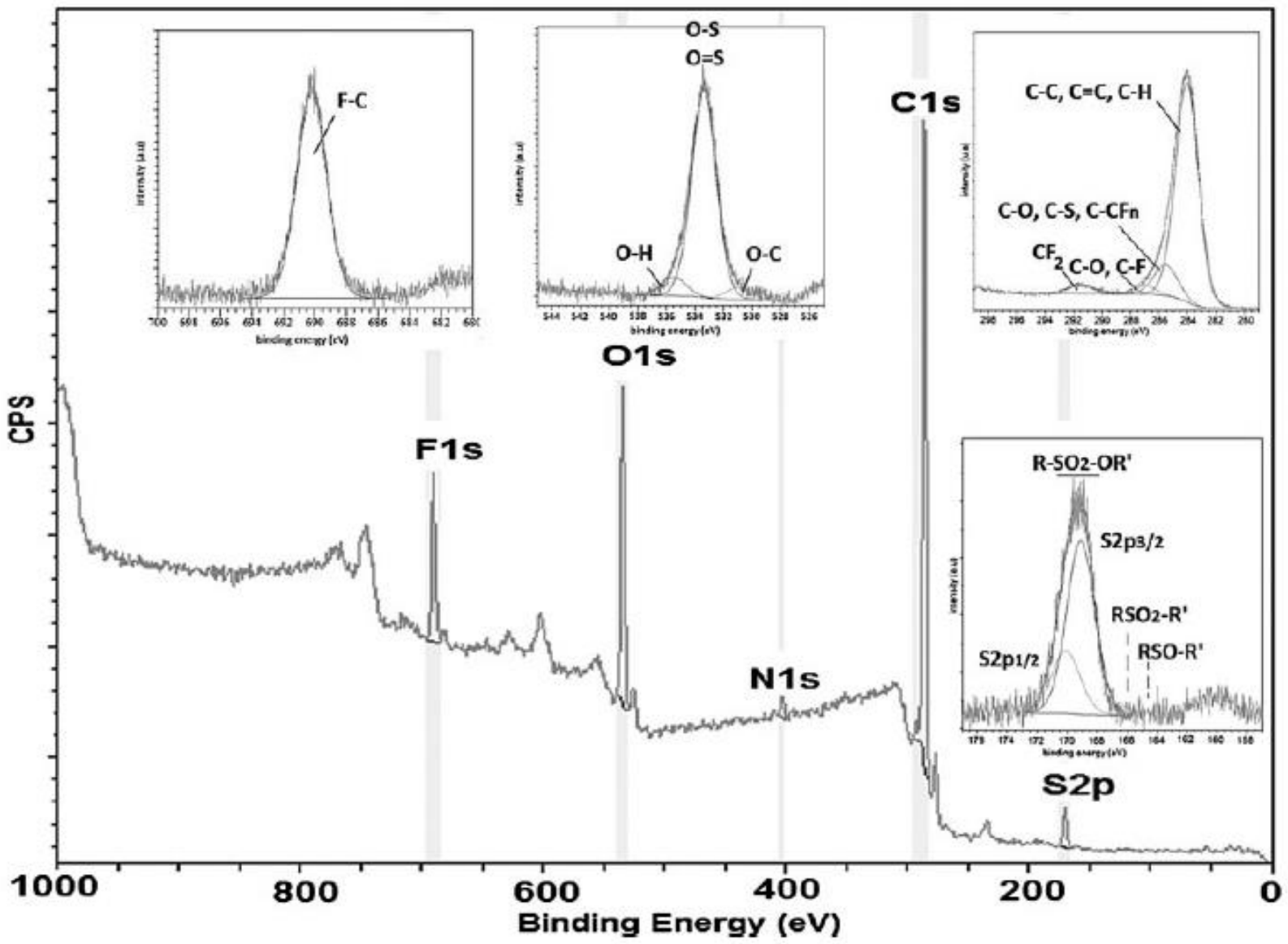

***Figure 3.** XPS survey with insets showing narrow scans corresponding to S2p, C1s, O1s, F1s of pp-sulfonated PS (He, 313 K for the both monomers, 1 kV).*

The significant retention of the sulfonic acids groups suggests milder plasma conditions at sub-atmospheric pressure in our synthesis conditions. This better retention could be explained by lower mean free paths of the radicals, excited species (such as metastables), and charged species. Their collisions frequency is therefore higher but their energy could be too low to fragment the sulfonic groups.

In this previous paper, the XPS results showed that the content in sulfonated groups was tunable by varying the temperature of the acid monomer (so the acid-to-styrene ratio) or by increasing the voltage.[23] Therefore, it was possible to produce plasma-membranes, thermally stable, with a higher content in sulfonic acids groups (up to 4% at 1 kV and up to 6.6% at 2 kV) than in the Nafion 117 (for which sulfonated groups of around 1% is reported). The presence of sulfonic acid groups was also confirmed by FTIR and static SIMS.[23]

The thermal stability was investigated by XPS from the point of view of the chemical composition and of the core level peak shape for the different elements of the membranes heated at 393 K in dry medium (stove) and at 353 K in aqueous medium (water bath). Owing to the similar elemental composition and the similar shape of the high resolution peaks before and after the heating (except for the oxygen peak after soaking in hot water, for which the OH/$H_2O$ component slightly increased), we can conclude that the thermal stability (from the point of view of the chemical composition) is correct. In an attempt to bring out the presence of easily accessible ionic sites, preliminary tests were operated by means of XPS on the pp-sulfonated PS immerged in an aqueous solution. They were soaked in NaOH 0.1 M during 30 min to replace the $H^+$ of the acid form by $Na^+$ in order to check if our films have a similar behavior to ionic exchange membranes. Afterwards, those ones were rinsed by milli-Q water, dried, and analyzed by XPS. The emergence of the Na1s peak at 1 075 eV and the Auger peak at 264 eV for the pp-sulfonated PS are clearly noticeable in Figure 4. The same treatment was applied to the non-sulfonated pp-PS in order to compare and confirm the ion exchange effect of the sulfonated polymer. The elemental composition of the reference (pp-PS) and pp-sulfonated PS after immersion in NaOH is given in Table 1.

Analogous tests were performed by Ogumi on pp-membranes synthesized at low pressure from trifluorochloroethylene and trifluoromethanesulfonic acid after soaking in CsOH during 30 min. The presence of Cs was attested by Electron Probe Micro-Analysis (EPMA) afterwards.[6]

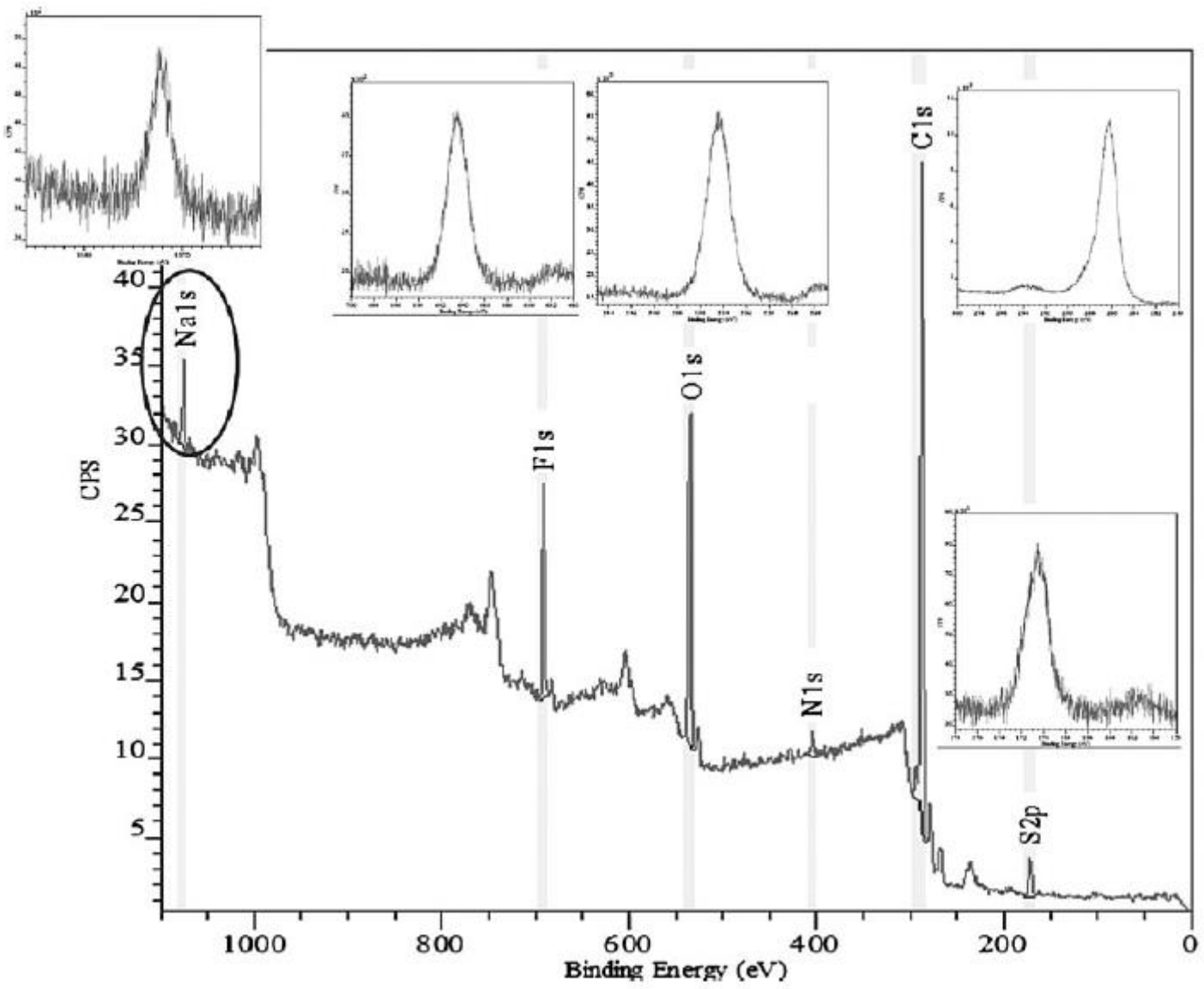

***Figure 4. Survey spectra after immersion in NaOH of pp-sulfonated PS with insets showing the narrows scans of S2p, C1s, O1s, F1s, Na1s.***

| Sample | Voltage [kV] | Si [%] | S [%] | C [%] | N [%] | O [%] | F [%] | Na [%] |
|---|---|---|---|---|---|---|---|---|
| pp-PS | 1.0 | 1.5 | 0 | 85.5 | 0 | 12.7 | 0 | 0.3 |
| pp-sulfonated PS/He | 1.0 | 2.0 | 2.9 | 72.3 | 1.0 | 15.9 | 4.9 | 1.0 |
| pp-sulfonated PS/He | 1.5 | 1.4 | 2.6 | 76.2 | 0.8 | 13.7 | 4.4 | 1.0 |
| RMS [%] | | 2 | <0.5 | 5 | 1 | 3 | 1 | <0.5 |

***Table 1. XPS elemental composition of pp-PS and pp-sulfonated PS after immersion in NaOH.***

**Deposition Rates**

Figure 5 shows a typical SEM cross-section image of pp-sulfonated PS synthesized onto silicon wafers. The films (few mm in thickness) are uniform and present a strong adhesion to the substrate whatever its nature. The deposition rates are given in Table 2. An increase in the deposition rate with the voltage is highlighted (360–580nm/min at 1 kV, against 825–1425nm/min at 1.5 kV) owing to an increase of the fragmentation and the amount of activated precursors. Therefore, the polymerization by random radical recombination is increased. Moreover, the films synthesized in the presence of Ar instead of He show a slightly higher deposition rate due to the more filamentary behavior of the Ar discharge.

| Carrier gas | Voltage [kV] | Polymerization rate [nm · min$^{-1}$] |
|---|---|---|
| He | 1 | 260–280 |
| Ar | 1 | 360–580 |
| Ar | 1.5 | 825–1 425 |

***Table 2. Deposition rates of the pp-sulfonated PS deposited onto Si wafers in the presence of Ar or He (313 K for the both monomers).***

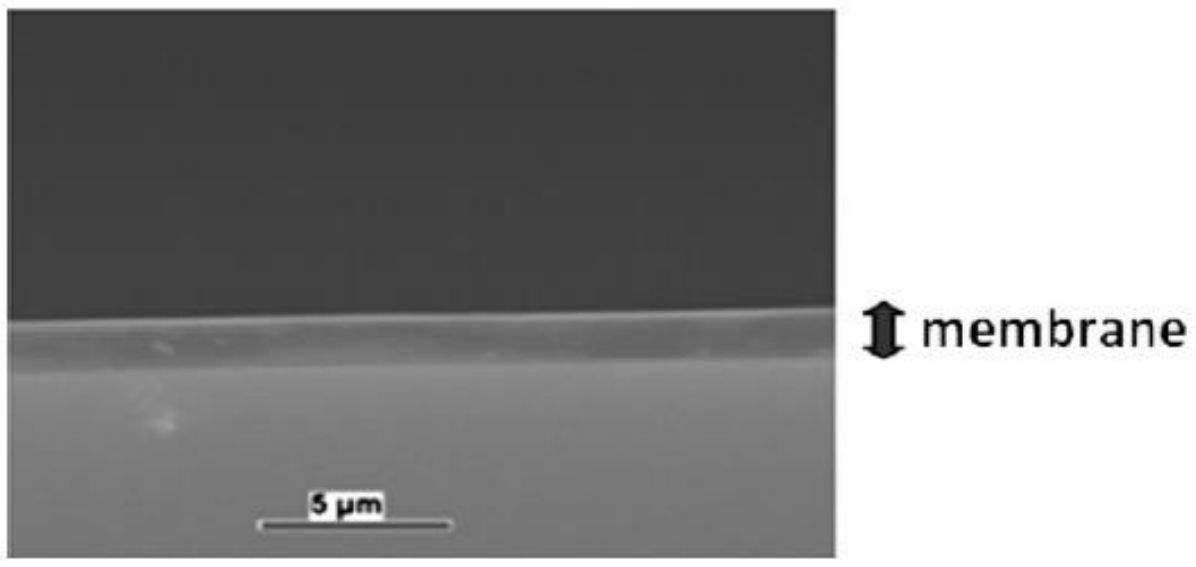

***Figure 5. SEM cross-section image of a pp-sulfonated PS on Si wafer.***

## 3.2. Nanoparticles Grafting on Carbon Substrates

**Chemical Structure**

The configuration illustrated in Figure 1 displays a good compromise between a high amount of Pt grafted and a low oxidation of the Pt nanoparticles. After any plasma treatment (with or without $O_2$), Pt nanoparticles were detected by XPS on the carbon surface; the fitted Pt and C high resolution peaks are presented in Figure 6 and 7, respectively.

Table 3 exhibits the elemental compositions of the metallic grafting on glassy carbon after ultrasonication and also on Carbon Toray Paper before and after ultrasonication. For comparison, the table indicates the composition of Carbon Toray Paper used as a substrate for the decoration of Pt nanoparticles in a pure argon discharge. Due to its significant porosity (78%), the amount of Pt detected on Carbon Toray Paper is lower than on glassy carbon. As the Carbon Toray Paper is a non-flat surface, the Pt nanoparticles are grafted over an indeterminate thickness, supposed to be higher than the depth probed by XPS analysis (10 nm).

| Substrate | Gas composition | Sonication | Pt [%] | Si [%] | C [%] | N [%] | O [%] |
|---|---|---|---|---|---|---|---|
| GC | $Ar/O_2$ | After | 4.3 | 3.8 | 67.7 | 5.5 | 18.7 |
| CT | $Ar/O_2$ | Before | 2.4 | 12.2 | 54.4 | 4.5 | 26.5 |
| CT | $Ar/O_2$ | After | 2.7 | 11.2 | 57.9 | 3.9 | 24.3 |
| CT | Ar | Before | 1.1 | 1.0 | 90.2 | 1.2 | 6.5 |
| CT | Ar | After | 0.7 | 1.1 | 87.9 | 1.2 | 9.1 |
| RMS [%] | | | 1 | 2 | 2 | <0.5 | 3 |

***Table 3. XPS elemental composition before and after ultrasonication of the surface of glassy carbon (GC) and Carbon Toray Paper (CT) after grafting of Pt nanoparticles by means of the atmospheric plasma torch.***

The incorporated oxygen is mostly due to the addition of oxygen gas (20mL.min$^{-1}$) into the Ar post-discharge (30 L.min$^{-1}$), to the opened environment in which the RF torch is working and to the colloidal solution. The silicon contamination (evidenced by XPS) may be due to the sprayer system. The small amount of nitrogen could be due to the capping agent of the Pt nanoparticles (PVP) and to the ambient air. The benefit of adding oxygen in the plasma was pointed out in previous publications.[15,16] As can be seen in Table 3, the amount of nanoparticles grafted is lower when no oxygen was added (less active sites) and the bonds become weaker (the quantity of Pt decreased after ultrasonication by a factor close to 2).

As the amount of platinum (when oxygen is added) was similar before and after ultrasonication, a strong adhesion of the nanoparticles to the carbon substrates is suggested. The activation of carbon substrates was argued in the Nano2hybrids project (especially by the way of simulation of grafting Au on carbon nanotubes).[17,31] The nucleation takes place preferentially on ‘‘defects’’ or on active sites (such as oxygenated vacancies) generated by plasma preactivation. Those ones trap metallic atoms and play the role of nucleation centers. Therefore, the density of defects influences the dispersion and the concentration of metal nanoparticles grafted and allows to adapt the interfacial properties. Therefore, thanks to the plasma activation, the distribution of the oxygenated vacancies is more homogeneous and leads to the grafting of a tunable concentration of nanoparticles.

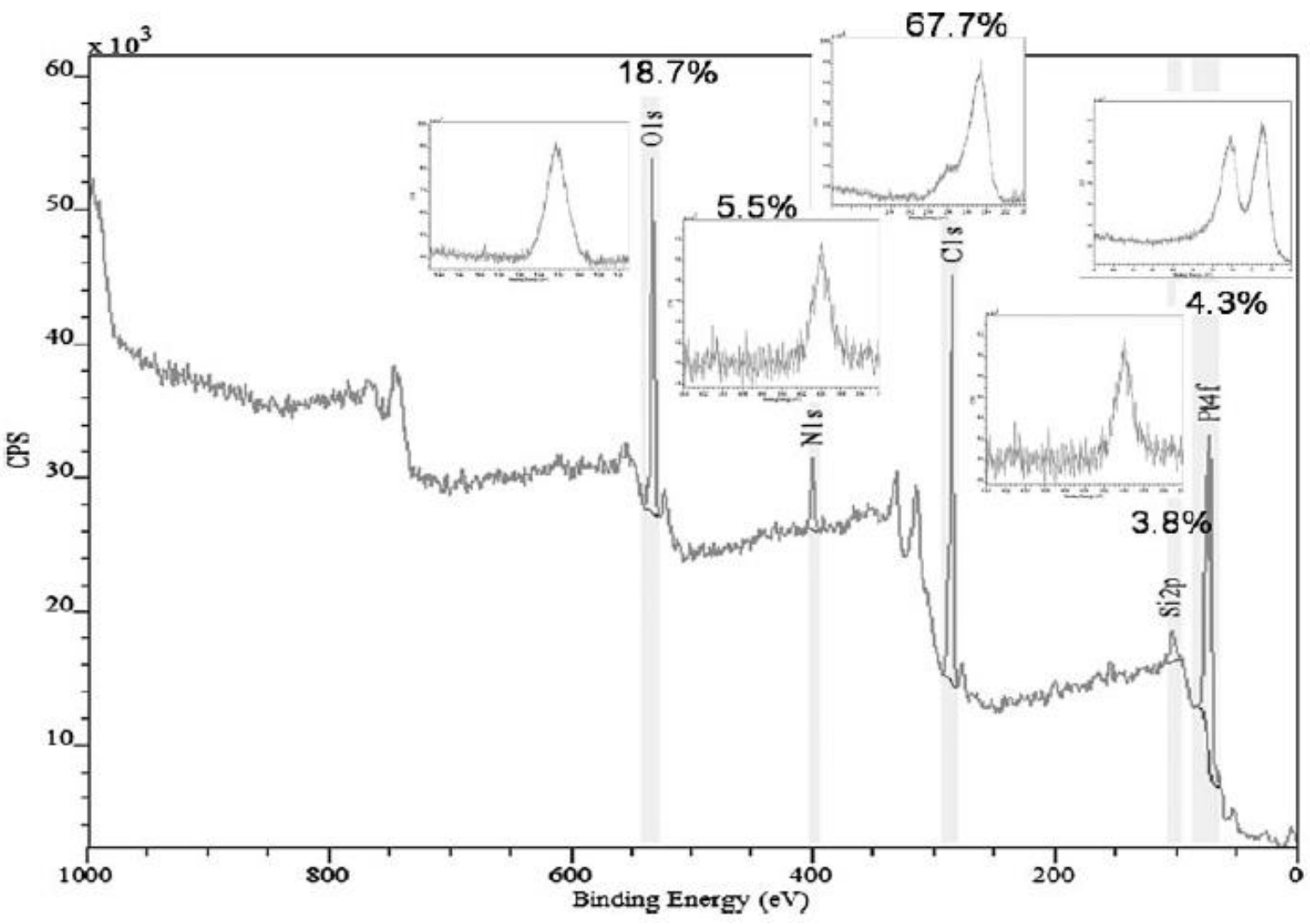

***Figure 6. XPS survey of Pt grafted on glassy carbon after ultrasonication with insets showing narrow scans of the elements.***

In Figure 7a, the fitting of the Pt 4f core level ($Pt4f_{5/2}$–$Pt4f_{7/2}$ spin-orbit doublets with 3.3 eV energy separation) was based on a method suggested by Croy.[32] The fitting for our deposits showed that the Pt was mostly present under metallic state for this configuration: about 70% of Pt (0), about 20% of Pt (+2), and less than 10% of Pt (+4). We should mention also that we have tested other plasma-spray-substrate configurations. Other configurations allow us to obtain more oxidized Pt. Depending on the geometry, the oxidation state of the platinum is tunable, which could play a role on the methanol oxidation reaction catalysis in the cell.[32] The fitting of the C1s peak is based on a method proposed by Swaraj.[33] The C1s high resolution scans evidenced the presence of carbon-oxygen species. As we can see in Figure 7b, the C spectrum was fitted in four components. C-O species are detected since during the plasma treatment, the sample was exposed to a strong oxidizing atmosphere including the ambient air, the injection of methanol droplets, and the addition of oxygen to the Ar gas.

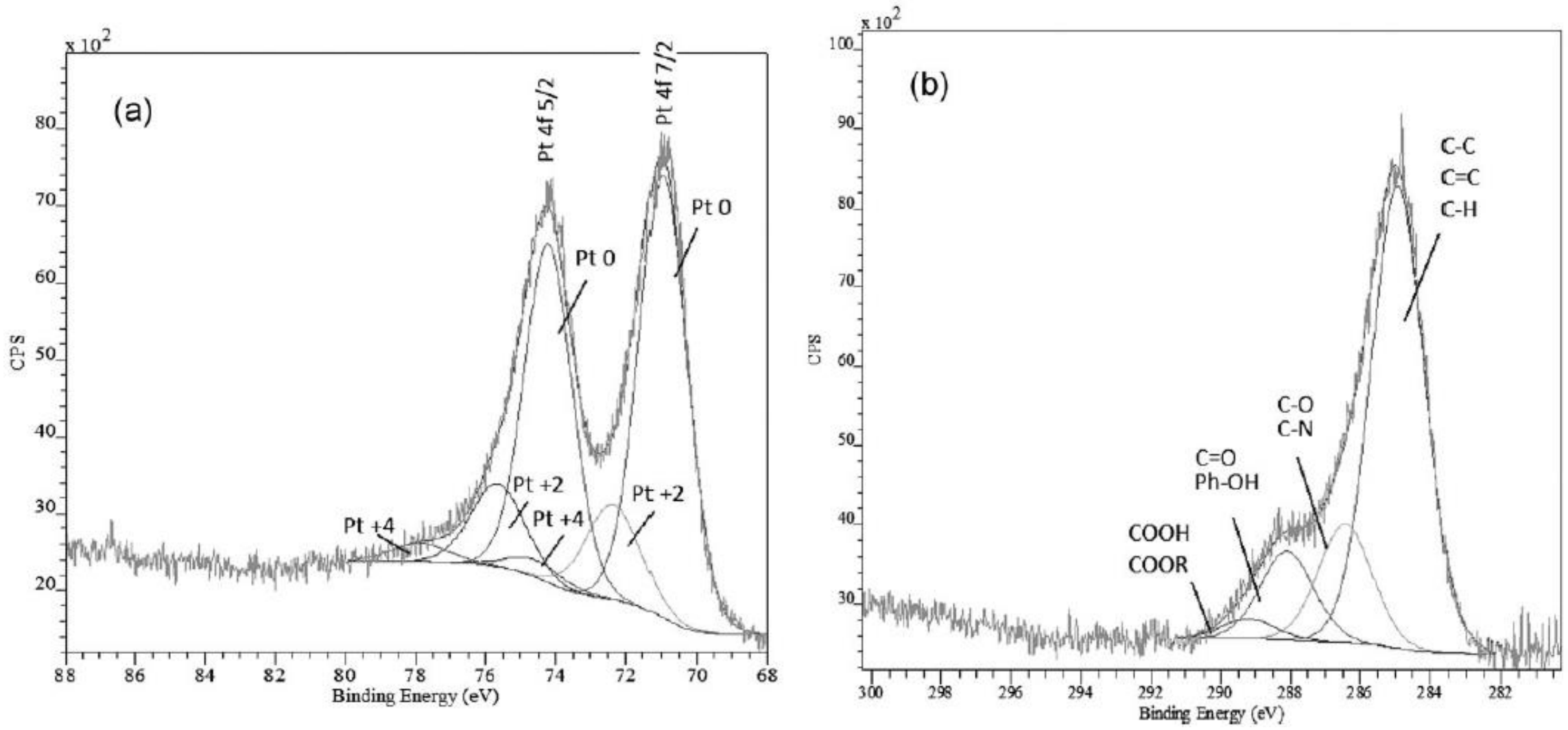

***Figure 7. Decomposition of the Pt4f (a) and C1s (b) XPS peaks for a grafting of Pt nanoparticles on glassy carbon after ultrasonication.***

The topmost surface of glassy carbon and Carbon Toray Paper decorated by nanoparticles was characterized by static SIMS. Figure 8 displays the positive and the negative Static SIMS spectra of glassy carbon decorated with Pt nanoparticles (in the presence of oxygen added in the postdischarge). The positive spectrum shows the existence of the following peaks: $Pt^+$, $PtO^+$, $PtO_2^+$, $PtCl_2^+$,, $PtCl_4^+$, $Pt_2^+$, $Pt_2O_x^+$, $Pt_3^+$, $Pt_3O_x^+$, $CH_3NH_3^+$, $NH_4^+$ (after grafting of the metallic particles), in addition to other peaks present on the surface of the native sample like aliphatics,$C_xH_yO^+$,$H_2O^+$,$Na^+$,$Al^+$, $Si^+$, $CO^+$, $NO^+$, $K^+$, $Ar^+$, and $Ca^+$, as confirmed in the literature.[34] The chlorine (non-visible in XPS spectra) could originate from the synthesis method of the nanoparticles. The nitrogen compounds come from the PVP used as a capping agent and from the ambient air. In the negative spectrum, peaks of $C^-$, $CH^-$, $OH^-$, $C_2H^-$, $CN^-$, $Cl^-$, $CNO^-$, $CO_2H^-$, $C_4H^-$, $SO_2^-$ , $SO_3^-$, $HSO_4^-$ (already present in the native sample) but also peaks of $PO_3^-$ , $PtO^-$, $PtO_2^-$ , $PtCl^-$, $PtCl_2^-$, $Pt_2O_x^-$, $Pt_3O_x^-$ , etc. are noticeable. The most abundant isotopes of platinum are $^{195}Pt$ (relative abundance: 0.338), $^{194}Pt$ (0.330), $^{196}Pt$ (0.252), and $^{198}Pt$ (0.072).

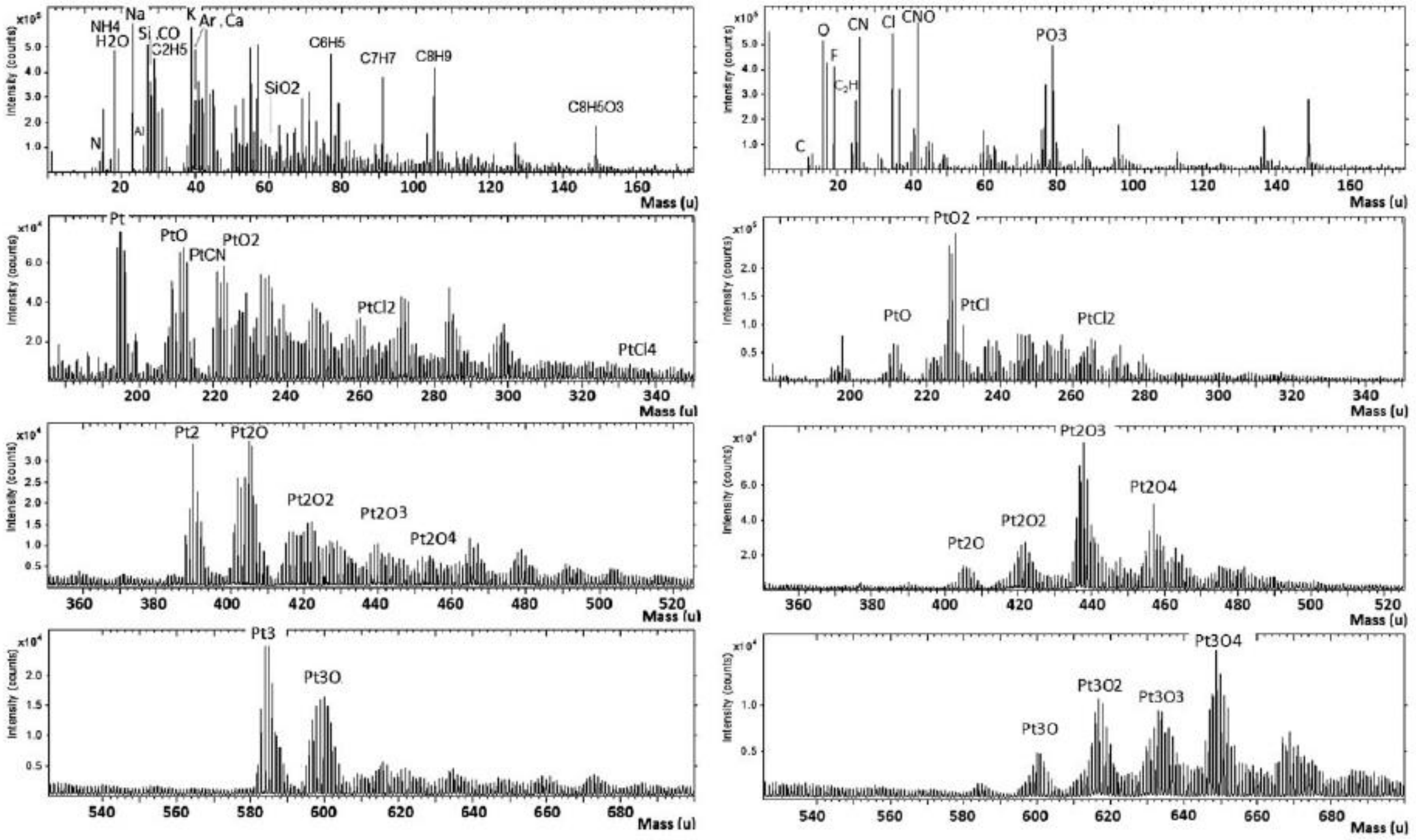

***Figure 8. Positive (left) and negative (right) static ToF-SIMS spectra of glassy carbon decorated with Pt nanoparticles.***

**<u>Morphology (SEM)</u>**

The SEM pictures shown in Figure 9 correspond to the glassy carbon substrate (a) as received and (b) decorated by Pt nanoclusters by the means of the RF plasma torch. In Figure 9b, an increase in the roughness is observed due to the grafting of the nanoclusters, which appear as white dots uniformly distributed. As a high loading leads to a more important coalescence process, the structure of the deposit is strongly related to the loading in Pt nanoparticles at the glassy carbon surface.

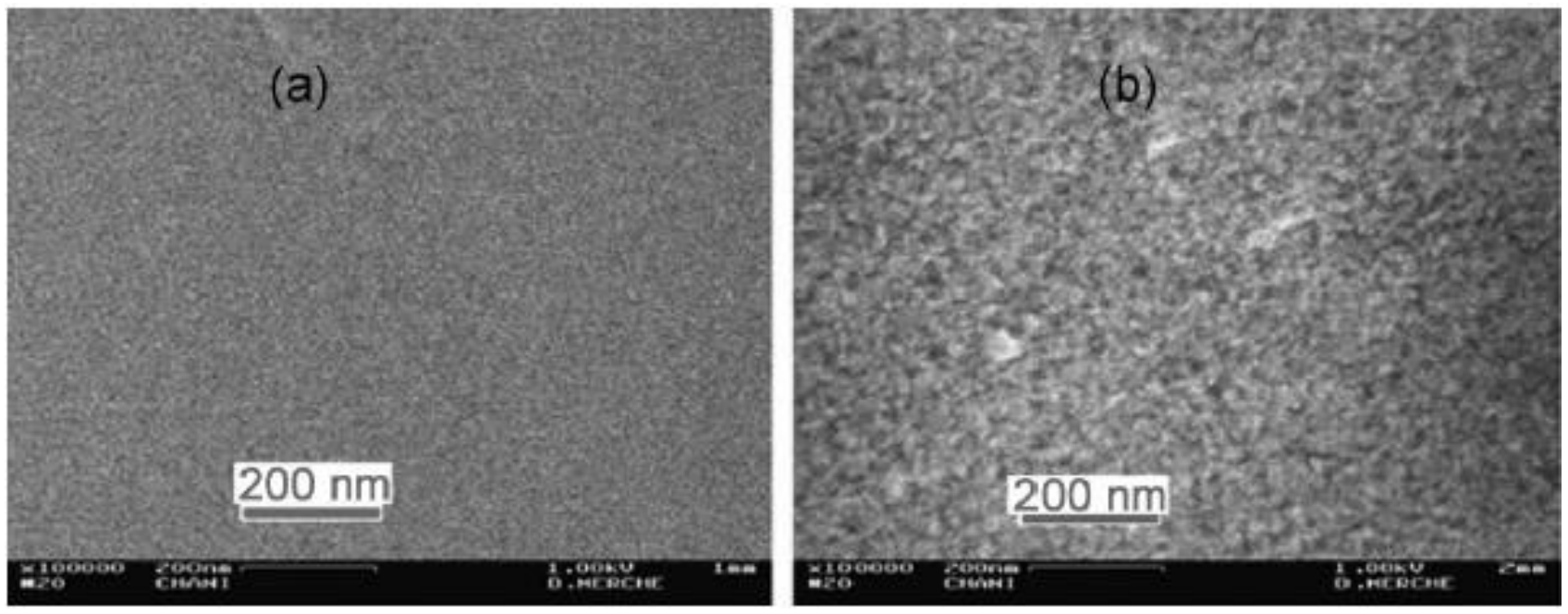

***Figure 9. SEM images of (a) glassy carbon untreated, (b) glassy carbon after grafting of catalyst.***

## 3.3. pp-Sulfonated PS on Electrodes

**Chemical Structure**

The pp-sulfonated PS synthesized on Pt nanoparticles deposited on carbon substrates were chemically characterized by XPS and by static ToF-SIMS. As those surface characterization results were similar to the previous ones operated on silicon wafers, they have not been repeated here. In order to determine the bulk homogeneity of the membranes and to have access to the membrane–catalyst and catalyst–carbon interfaces, dynamic ToF-SIMS depth profiling has been performed. The evolution of the dynamic ToF-SIMS negative fragments as a function of the erosion time for the pp-sulfonated membranes on glassy carbon loaded with Pt clusters is shown in Figure 10.

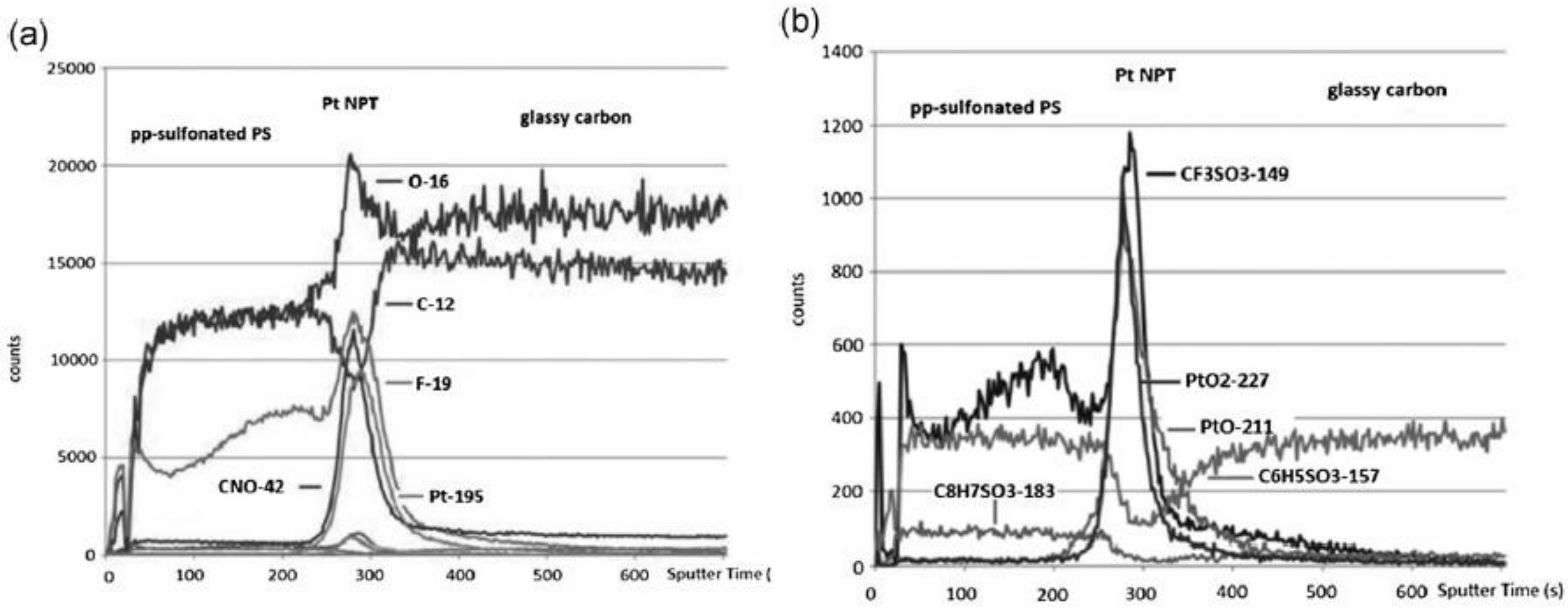

***Figure 10. (a) Evolution of the negative dynamic ToF-SIMS fragments from pp-sulfonated PS (DBD Ar, 313 K for the two monomers, 1 kV, 3 min) deposited on glassy carbon decorated with Pt nanoparticles versus the time of erosion; (b) zoom of (a) for the low intensity fragments.***

Figure 10 shows that the sulfonic acid groups were homogeneously grafted inside the whole membrane (fragments at m/z=157 and 183 corresponding to $C_6H_5SO_3^-$ and $C_8H_7SO_3^-$, respectively). The non-fragmented trifluoromethanesulfonic acid monomer (m/z=149) is mostly adsorbed at the ‘‘plasma-membrane’’/‘‘Pt nanoparticles’’ interface due to the two monomers injected before the breakdown of the plasma in the DBD chamber and the polymerization process. The fluorine coming from the fragmentation of the trifluoromethanesulfonic acid follows the same trend. The polymeric film/Pt and Pt/glassy carbon interfaces are particularly visible; the carbon components without nitrogen decrease during the erosion of the nanoparticles and increase when the Pt nanoparticles have totally disappeared. The $CNO^-$ peak at m/z=42 (from the capping agent of the colloidal solution) and oxygen (after the achievement of a plateau in the plasma-film) increase similarly to the Pt. The oxygen bond to the Pt comes from the tiny flow of oxygen added to the Ar plasma, from the solution and from the open environment.

**Morphology (SEM)**

Figure 11a shows the SEM pictures of the top view of scratched pp-sulfonated PS synthesized on glassy carbon loaded in metallic catalyst. The plasma-polymer is flat and homogeneous. Figure 11b shows the cross-section of the same deposit. Pt clusters, homogenously distributed and clearly visible at the interface (as confirmed by the dynamic ToF-SIMS results), appear bright (effect of the atomic number).

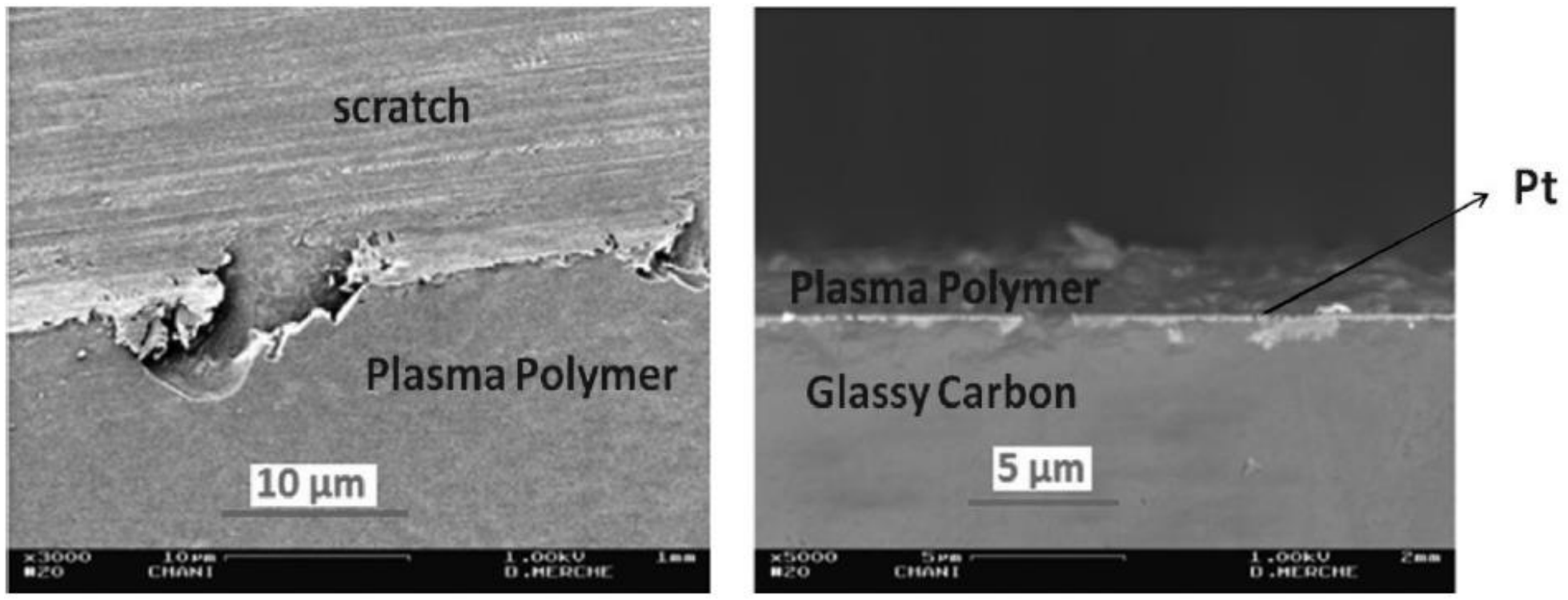

***Figure 11. SEM images of pp-sulfonated PS synthesized on glassy carbon loaded with Pt nanoparticles.***

# 4. Conclusion

We tried to show, in this paper, that atmospheric pressure plasma is a potential useful tool for building membrane-catalyst-electrode assemblies for fuel cell applications. It presents the advantage of being an easy-to-set-up process that does not require complicated technology or involve complex chemical reaction schemes.

Thermally stable pp-sulfonated PS coatings have been successfully synthesized by means of a home-built DBD, in one single step, by ''copolymerization'' of two precursors (styrene and trifluoromethanesulfonic acid) injected simultaneously into the discharge. These coatings could play the role of ion-exchange membranes in miniaturized PEMFC, using hydrogen or methanol for portable applications. The complementarity of the spectroscopic techniques (XPS, FTIR, static and dynamic SIMS) showed that the sulfonic acid groups were well preserved in the discharge near atmospheric pressure and also grafted inside the whole polystyrene membrane. The influence of the parameters (temperature of the acid, voltage applied between the electrodes, nature of the carrier gas, etc.) on the amount of the grafted ionizable groups was monitored by XPS, which showed a higher content than in the Nafion 117 (up to4% at 1 kV and up to 6.6% at 2 kV versus about 1% for the Nafion). The presence of ionic sites was also pointed out by means of the Na signal followed by XPS after soaking the membranes in a solution of NaOH to change its form.

In order to develop membrane-electrode assembly by atmospheric plasma, these films were also deposited directly onto carbon substrates decorated with platinum. We have grafted the catalyst from a platinum colloidal solution sprayed into the post-discharge of an RF atmospheric plasma torch on gas diffusion layers and on glassy carbon (used as a model for the depth profiling by dynamic SIMS) in several configurations and for different parameters. The amount of Pt grafted on carbon substrates and the oxidation states of the Pt can be tuned depending on the configurations. This one-step process is a promising technique to deposit nanocatalysts for fuel cells applications due to its easy transferability to industrial process.

The aim of this process was to constitute electrodes that will be used afterwards as substrates for the adhesion of plasma-membranes as part of membrane-electrode assemblies for fuel cells. Thedifferent interfaces of themembraneelectrode assembly are clearly visible, detected, and characterized by depth profiling (dynamic ToF-SIMS) and SEM.

# 5. Acknowledgements

This work was supported by the Belgian Federal Government IAP (Belgian Interuniversity Attraction Pole ‘‘Physical Chemistry of Plasma-Surface Interactions – project P6/08) and by the FNRS (FRFC grant n 2.4543.04). D. Merche thanks the F. R. I. A. (Belgium) and the ‘‘Fondation Jaumotte – Demoulin-Van Buuren’’ for a PhD grant.